\documentclass[aps,preprint,amsmath,amssymb,amsfonts,nofootinbib]{revtex4}
\usepackage{epsfig}
\usepackage{graphicx}
\usepackage{dcolumn}
\usepackage{bm}
\usepackage{amsthm}
\usepackage{amsmath}
\usepackage{color}
\usepackage{setspace}

\newcommand{\beq}{\begin{equation}}
\newcommand{\eeq}{\end{equation}}
\newcommand{\bea}{\begin{eqnarray}}
\newcommand{\eea}{\end{eqnarray}}

\begin{document}

\vskip 1cm

\title{Entanglement and Thermal Entropy of Gauge Fields}
\author{Christopher Eling$^1$}
\author{Yaron Oz$^2$}
\author{Stefan Theisen$^1$}

\affiliation{$^1$ Max Planck Institute for Gravitational Physics,
Albert Einstein Institute, Potsdam 14476, Germany}
\affiliation{$^2$ Raymond and Beverly Sackler School of Physics and Astronomy,
Tel Aviv University, Tel Aviv 69978, Israel}

\date{\today}

\begin{abstract}

We consider the universal logarithmic divergent term in the entanglement entropy
of gauge fields in the Minkowski vacuum with an entangling sphere.
Employing the mapping in arXiv:1102.0440, we analyze the corresponding thermal
entropy on open Einstein universe
and on the static patch of de Sitter. Using the heat kernel of the vector
Laplacian we resolve a discrepancy between the free field calculation and the expected
Euler conformal anomaly.
The resolution suggests a modification of the well known formulas for the
vacuum expectation value of the spin-1 energy-momentum tensor
on conformally flat space-times.

\end{abstract}


\maketitle



\section{Introduction and Summary}

Consider a quantum mechanical system in its pure ground state $|0\rangle$,
and let us suppose that it is
divided into two subsystems $A$ and $B$. Accordingly, we write the total
Hilbert space of the system as a direct product of the two subspaces
${\cal H}_{tot} = {\cal H}_A \otimes {\cal H}_B$.
The entanglement entropy $S_{A}$ of the subsystem $A$ is the von Neumann entropy
\begin{align}
S_{A} = -{\rm Tr} \rho_A \ln \rho_A \  ,
\label{EE}
\end{align}
where $\rho_A$ is the reduced  density matrix of the subsystem $A$, obtained by tracing the total density matrix of the system over the Hilbert subspace
${\cal H}_B$
\begin{align}
\rho_A={\rm Tr}_{B} |0\rangle \langle 0|  \  .
\label{reduced}
\end{align}

A difficulty arises when applying this to compute the entanglement
entropy of gauge fields. One way to see this is
in the canonical quantization framework, where the total system is on a time slice,
with $A$ and $B$ denoting now two complementary spatial
regions. The Hilbert space of states of the system is defined modulo
gauge transformations, and it does not have an obvious
factorization as a direct product of the Hilbert spaces of the two subsystems.
Thus, there is a subtlety in defining in a gauge invariant way
the reduced density matrix (\ref{reduced}) and the entanglement entropy (\ref{EE}).
The same issue arises in the path integral formalism.
A proper resolution is likely to require an understanding of the
role played by the gauge fields on the entangling surface that divides $A$ from $B$.

In this letter we will address a related issue when mapping the entanglement entropy
of the gauge fields to a thermal entropy.
The gauge field theories that we will discuss are conformal field theories (CFTs).
Consider the flat four-dimensional Minkowski space-time $R^{1,3}$,
where the entangling surface is a two-sphere with radius $\mathcal{R}$, and
we trace over the gauge field degrees of freedom in the spatial volume
outside the spherical region (subsystem $B$).
It has been shown in \cite{Casini:2011kv} that the causal
development of the ball (subsystem $A$) is in the same conformal class as
the open Einstein universe $R \times H^3$ ($H$ is hyperbolic space) and the
static patch of de Sitter space (dS).
The maps are explicitly constructed in
\cite{Casini:2011kv} and its is shown that
they map the entanglement entropy of the sphere to a thermal entropy
on $R\times H^3$ and the static patch of dS.
Note, that both the open Einstein universe and the static patch of
de Sitter are conformally related to Rindler space, therefore these
thermal states are all related to each other by conformal transformations
\cite{Candelas:1978gf}.

Both the entanglement and the corresponding thermal entropies are UV divergent
and have a logarithmically divergent term whose coefficient is
regularization independent,
\begin{align}
S_{\rm log} = 4\,a\, \ln(\delta/\mathcal{R}) \ . \label{logterm0}
\end{align}
$\delta$ is the UV cutoff and $a$ is the Euler conformal anomaly coefficient.
This universal term in the entanglement entropy can be seen by using the replica trick
\cite{CallanWilczek,Ryu:2006ef}
\begin{align}
\delta \frac{\partial S}{\partial \delta} =  \lim_{n \rightarrow 1}
\frac{\partial}{\partial n} \langle \int \sqrt{-g}\, T^\mu_\mu \rangle_{M_n}\,.
\end{align}
$M_n$ is an $n$-sheeted manifold, and the
conformal anomaly in four dimensions
\begin{align}
\langle T^\mu_\mu\rangle_g = -\frac{a}{16\pi^2} E_4 + \frac{c}{16\pi^2} C^2 \  .
\end{align}
Here $\sqrt{g}E_4$ with
\begin{equation}
E_4=R_{\mu\nu\rho\sigma}R^{\mu\nu\rho\sigma}-4\,R_{\mu\nu}R^{\mu\nu}+R^2
\end{equation}
is the Euler density and
\begin{equation}
C^2=R_{\mu\nu\rho\sigma}R^{\mu\nu\rho\sigma}-2\,R_{\mu\nu}R^{\mu\nu}+{1\over3}R^2
\end{equation}
the square of the Weyl tensor. For
the spherical entangling surface one gets (\ref{logterm0}).

While the free field thermal entropy calculations for spin-0 and spin-1/2
reproduce correctly (\ref{logterm0}), a mismatch for spin-1 fields
has been noticed on static de Sitter in \cite{Dowker:2010bu}.
Indeed, we will see a similar mismatch on open Einstein space-time.
Using the heat kernel of the vector Laplacian we will resolve the mismatch
and will discuss the relation between the
discrepancy  and the surface term introduced by Kabat \cite{Kabat:1995eq}.
The resolution suggests a modification to the well known formulas for the
vector contribution to the vacuum expectation value of the energy-momentum tensor
on conformally flat space-times.
We note that for the thermal entropy of $\mathcal{N}=4$ SYM on the open Einstein universe
our calculations imply that the thermal entropy at  temperature
$(2\pi \mathcal{R})^{-1}$ is the same at weak and strong
coupling, and is not larger at strong coupling contrary to the result of
\cite{Emparan:1999gf}.

The paper is organized as follows. In Section 2 we briefly review the map of
\cite{Casini:2011kv} from entanglement entropy to thermal entropy.
We calculate the logarithmically divergent term in thermal entropy of free fields on
open Einstein space-time using a well known formula for the expectation value
of the stress-energy tensor and show a mismatch with the expected result.
We then resolve the discrepancy by a heat kernel analysis of the Laplacian.
In Section 3 we consider the thermal CFT at strong coupling
and show that no discrepancy similar to the one at weak coupling arises. In Section 4 we
propose the modification to
the formulas in the literature  for the vacuum expectation value of the energy-momentum
tensor
on conformally flat space-times. We argue that the modification is due to the surface term
proposed by Kabat.

\section{CFT Entanglement and Thermal Entropy}

\subsection{From Entanglement to Thermal Entropy}

We consider the case when the entangling surface is a sphere $S^{d-2}$ in flat space-time
$R^{1,d-1}$, and one traces out the subsystem $B$, corresponding to the $(d-1)$-dimensional
volume outside the
entangling sphere. We write the flat metric in polar coordinates
\begin{align}
ds^2 = -dt^2 + dr^2 + r^2 d\Omega^2_{d-2} \label{sphere} \ ,
\end{align}
where $d\Omega^2_{d-2}$ is the metric on the unit sphere. The entangling
surface is defined as $t=0$, $r=\mathcal{R}$ ($\mathcal{R}$ is the radius of the sphere). As
discussed in \cite{Casini:2011kv}, there are two different sets of coordinate
transformations which map the causal
development of the volume inside the entangling sphere in flat space-time into either the
open Einstein universe $R \times H^{d-1}$ or the static patch of de Sitter (dS). Both
examples will be useful for our discussion.

First, one can make the coordinate transformation from $(t,r)$ to new coordinates $(\tau,u)$
\begin{align}
t=& \mathcal{R} \frac{\sinh(\tau/\mathcal{R})}{\cosh u + \cosh(\tau/\mathcal{R})}\,,\qquad
r=\mathcal{R} \frac{\sinh u}{\cosh u + \cosh(\tau/\mathcal{R})} \label{map} \ .
\end{align}
As $\tau \rightarrow \pm \infty$, $t=\pm \mathcal{R}$ and $r=0$, while for
$u \rightarrow \infty$, $t=0$ and $r=\mathcal{R}$. The new coordinates cover the
causal development of the spherical region $r \leq \mathcal{R}$.
In these coordinates the metric is conformal to $R \times H^{d-1}$
\begin{align}
ds^2 = \Omega^2(u, \tau) \left(-d\tau^2 + \mathcal{R}^2(du^2 + \sinh^2 u
~d\Omega^2_{d-2})\right) \ .
\end{align}
It was proven in \cite{Casini:2011kv} that for a CFT
the entanglement entropy with the entangling
sphere is equivalent to a thermal entropy on the hyperbolic space at the
special temperature $T_0= (2\pi \mathcal{R})^{-1}$.
The identification of the  entanglement entropy with a thermal entropy follows
from the equivalence of the reduced density matrix
to $e^{-H/T_0}$ where $H$ is a local Hamiltonian. It is analogous to the
identification of the entanglement entropy of the half space $(t=0,x>0)$ in Minkowski
space-time, whose causal development is the Rindler wedge,
and the Rindler thermal entropy.

The second type of transformation is
\begin{align}
t=& \mathcal{R} \frac{\cos \theta \sinh(\hat{\tau}/\mathcal{R})}
{1+\cos \theta \cosh(\tau/\mathcal{R})}\,,\qquad
r =\mathcal{R} \frac{\sin \theta}{1+\cos \theta \cosh(\hat{\tau}/\mathcal{R})} \label{map}.
\end{align}
In this case, as $\hat{\tau} \rightarrow \pm \infty$, $t=\pm \mathcal{R}$ and $r=0$,
while for $\theta \rightarrow \pi/2$, $t=0$ and $r=\mathcal{R}$. Making the further
transformation $\hat{r} = \mathcal{R} \sin \theta$, the flat metric (\ref{sphere})
becomes the static patch of dS up to an overall conformal factor
\begin{align}
ds^2 = \hat\Omega^2(\hat{r}, \hat{\tau}) \left(-(1-\frac{\hat{r}^2}{\mathcal{R}^2})
d\hat{\tau}^2 + \frac{d\hat{r}^2}{1-\frac{\hat{r}^2}{\mathcal{R}^2}}+\hat{r}^2
d\Omega^2_{d-2} \right) \ .
\end{align}
The causal development of the spherical region is mapped into the region outside
the cosmological horizon at $\hat{r} = \mathcal{R}$. In this case the entanglement entropy
is equivalent to the thermal entropy of fields on static dS at the de Sitter temperature
$T_0 = (2\pi \mathcal{R})^{-1}$.

The thermal entropy density and the energy density are related via the first law of thermodynamics
\begin{align}
s(T_0) = \int^{T_0}_0 \frac{d\epsilon(T)}{T} \label{firstlaw} \ .
\end{align}
The total thermal entropies
on open Einstein $S_{\rm oE}$ and on static dS $S_{\rm dS}$  are  divergent, as expected
from their equivalence to the entanglement entropy. In the open Einstein case, the divergence arises due to the integration over the infinite spatial volume. On the static patch of dS, the spatial volume is finite
but the entropy density diverges in the vicinity of the horizon. To render the volume integrals finite, we introduce a UV cutoff $\delta$.
In the original coordinates (\ref{sphere})
we cut off the integration at a distance $\delta$ from $\mathcal{R}$, i.e.
$r_{\rm max}={\cal R}-\delta$
with $\delta/\mathcal{R} \ll 1$.
In the hyperbolic coordinates, this translates into the condition
$u_{\rm max} = -\ln (\delta/2\mathcal{R})$ while on the static dS patch
$\hat r_{\rm max}={\cal R}-{1\over 2{\cal R}}\delta^2$.

\subsection{Thermal entropy of free fields on open Einstein space-time}
\label{freecalcs}

For the case of free field theories, the calculation of various thermodynamical quantities
on four-dimensional static space-times was performed many years ago
\cite{Candelas:1978gf,Brown:1986jy}.
For the open Einstein universe it was found
\begin{align}
\langle T_\mu^\nu \rangle_{\rm oE} = \sum_s \frac{n_s h(s)}{6\pi^2 \mathcal{R}^4}
\int^{\infty}_{0}d\lambda\frac{\lambda(\lambda^2+s^2)}{e^{\beta\lambda/\mathcal{R}}
-(-1)^{2s}} {\rm diag}(-3,1,1,1) \ .
\end{align}
$n_s$ is the number of fields of spin $s$ and $h(0)=1,
h(1/2)=h(1)=2$ the number of physical propagating degrees of freedom for
real conformally coupled scalars, Weyl fermions and gauge bosons, respectively.
Performing the integrals, one finds
\begin{align}
\langle T_\mu^\nu \rangle_{\rm oE} = \frac{\pi^2}{90\beta^4}
\left(n_0+\frac{7}{4}n_{1/2}+2\,n_1+\frac{5\beta^2}{8\pi^2\mathcal{R}^2}(n_{1/2}+8\,n_{1})
\right){\rm diag}(-3,1,1,1) \label{weakstress}.
\end{align}
To evaluate the entropy, we use  (\ref{firstlaw}) with
$\epsilon_{\rm oE}=T^0{}_0$ and get
\begin{align}
S_{\rm oE}^{\rm free} = \frac{2\pi^2 V_3}{45 \beta^3} \left(n_0 + \frac{7}{4} n_{1/2}
+ 2\,n_1 + \frac{15\beta^2}{16\pi^2 \mathcal{R}^2} (n_{1/2}+8\,n_1)\right) \ ,
\end{align}
where the regulated three-volume element on the hyperbolic space is
\begin{align}
V_3 =  \mathcal{R}^3 \int^{u_{\rm max}}_{0} \sinh^2 u\, du \int \sin\theta d\theta\,d\phi=
4\pi \mathcal{R}^3 \int^{u_{\rm max}}_{0} \sinh^2 u\, du.
\end{align}
With $\beta = 2\pi \mathcal{R}$ and expanding out $V_3$,
one finds the logarithmic divergent term\footnote{We do not display the
leading term which scales as ${\cal R}^2/\delta^2$ or the finite terms.}
\begin{equation}
S^{\rm free}_{\rm oE} = \frac{1}{90}\left(n_0 + {11\over2}n_{1/2} + 32\, n_1 \right)
\ln(\delta/\mathcal{R})  \ . \label{weakentropy}
\end{equation}
Thus, we find a mismatch for the spin-1 field with (\ref{logterm0}) since the
Euler anomaly coefficient reads
\begin{equation}\label{acoeff}
a = \frac{1}{360} \left(n_0 + {11\over2} n_{1/2} + 62\, n_1\right) \ ,
\end{equation}
This mismatch was also
noticed by Dowker \cite{Dowker:2010bu}, who performed essentially the
same calculation of thermal entropy for free fields, but on the static patch of dS.

One can relate the entropies in dS and oE by starting with the
observation \cite{BrownCassidy}
that the energy-momentum tensors of conformal field theories
on conformally related manifolds
$({\cal M},g_1)$ and $({\cal M},g_0)$ with $(g_1)_{\mu\nu}=e^{2\sigma(x)}(g_0)_{\mu\nu}$
and $g_0$ flat are related as
\begin{equation}\label{ee}
\sqrt{-g_1}\langle T^\mu{}_\nu\rangle_{g_1}=\sqrt{-g_0}\langle T^\mu{}_\nu\rangle_{g_0}
+\hbox{anomaly terms}\,.
\end{equation}
This relationship holds also at finite temperature for theories on static manifolds.
At non-zero temperature the anomaly term remains
temperature independent and simply shifts the zero-temperature
values of the energy density $\epsilon$. It therefore does not affect the calculation of the thermal entropy.
We will now use \eqref{ee} to relate the
entropies of the open Einstein and the de Sitter universes, which are
both conformal to flat Rindler space.

Taking the static dS metric, pulling out an overall factor,
\begin{align}
ds^2 =  (1-\frac{\hat{r}^2}{\mathcal{R}^2}) \left(-d\hat{\tau}^2 + \frac{d\hat{r}^2}{(1-\frac{\hat{r}^2}{\mathcal{R}^2})^2}+\frac{\hat{r}^2}{1-\frac{\hat{r}^2}{\mathcal{R}^2}}
d\Omega^2_{d-2} \right) \ ,
\end{align}
and making the coordinate transformation $r = \mathcal{R} \tanh u$, one finds
\begin{align}
ds^2_{\rm dS} = \frac{1}{\cosh^2 u} ds^2_{\rm oE} = \frac{\mathcal{R}^2}{\xi^2 \cosh^2 u} ds^2_{\rm Rindler},
\end{align}
where the metric for the flat Rindler space has the form\footnote{For the relationship between the Rindler $(\xi,x,y)$ and $(u,\theta, \phi)$ see the Appendix of \cite{Candelas:1978gf}.}
\begin{align}
ds^2_{\rm Rindler} = -\frac{\xi^2}{\mathcal{R}^2} dt^2 + d\xi^2 + dx^2 + dy^2  \  \label{Rindlermetric}.
\end{align}
Thus, up to temperature independent anomaly contributions, it follows from (\ref{ee}) that
\begin{align}
\sqrt{g}_{\rm dS} \epsilon_{\rm dS} = \sqrt{g}_{\rm oE} \epsilon_{\rm oE} = \sqrt{g}_{\rm Rindler} \epsilon_{\rm Rindler}. \label{conformale}
\end{align}

In the presence of a time-like Killing vector $\chi^\mu$ the conserved energy has the general form (cf. also the discussion in \cite{Casini:2011kv})
\begin{equation}
E=\int_{\Sigma} \langle T_{\mu\nu}\rangle\, \chi^\mu n^\nu d \Sigma,
\end{equation}
where the integral is over a space-like surface $\Sigma$, e.g. a constant time slice,
and $n^\mu$ is a unit normal. For the dS metric this becomes
\begin{align}
E_{\rm dS}&=\int \big(T_0^0\,\sqrt{g_{00}}\big)_{\rm dS}\,d \Sigma
=4\pi\,\int_0^{\hat r_{\rm max}}\!\!\!\!\!
(T_0^0)_{\rm oE}{\hat r^2\, d\hat r\over\big(1-(\hat r/{\cal R})^2\big)^2}\,.
\end{align}
In the last step we have used \eqref{conformale} and we have integrated
over the angular coordinates. As explained above we have dropped the temperature
independent contributions. The entropy follows from
\begin{equation}
S=\int_0^{1/(2\pi {\cal R})}{1\over T}{d\,E\over d\,T}dT\,.
\end{equation}
If we cut-off the radial integral at $\hat r_{\rm max}$ and extract
the log-divergence we find again \eqref{weakentropy}.

\subsection{Free fields on static de Sitter - heat kernel analysis}

We now consider the calculation of the entropy in a different form using the effective action on static dS. The one-loop effective action -- for free fields this is all there is --
can be computed via the heat kernel method. The method is well known and
the relevant results can be found in \cite{CD}. The effective action is
\begin{equation}
W=-{1\over 2}\,n_0\,\log\det\Delta_0^{(\xi=1/6)}+{1\over2}\,n_{1/2}\,\log\det\Delta_{1/2}
-{1\over2}\,n_1\,\big(\log\det\Delta_1-2\log\det\Delta_0^{(\xi=0)}\big)
\end{equation}
where $\Delta_s$ are the fluctuation operators. The contribution
from the gauge fields includes the ghosts.

The heat kernel expansion of the effective action follows from
\begin{equation}
\log\det\Delta=\int d^4 x\sqrt{g}\left({1\over2}\Lambda^4\, b_0+\Lambda^2\, b_2
+\log(\Lambda^2)\, b_4+\hbox{finite}\right)
\end{equation}
where we have specified to $d=4$.
$\Lambda$ is a UV momentum cut-off and
the $b_{2n}$ are
scalars composed of the curvature and covariant derivatives with
$2n$ derivatives of the metric. To extract the log-divergence we need
$b_4$.

The effective action $W$ is related to the free energy $F$ via
$W=\beta F$. Using the thermodynamic relation $S=-{\partial F\over\partial T}$
one obtains for the entropy $S$
\begin{equation}
S=(\beta\partial_\beta-1)W\,.
\end{equation}
We are interested in the entropy at the temperature $T_0=1/(2 \pi {\cal R})$.
To introduce an arbitrary temperature in the geometric expression for the effective action
one continues to Euclidean signature and restricts the angular coordinate in the
$(\hat r,\hat\tau)$ plane to
$(\hat\tau/{\cal R})\in[0,(\beta/{\cal R})]\equiv[0,2\pi\alpha]$.
For $\beta\neq 2\pi{\cal R}$ this introduces a conical singularity at the origin of the
$(\hat r,\hat\tau)$ plane with deficit angle $2\pi(1-\alpha)$.
We therefore need to compute
\begin{equation}\label{EntropydS}
S=(\alpha\partial_\alpha-1)W(\alpha)\big|_{\alpha=1}
\end{equation}
We will denote by $\Sigma$ the codimension two surface containing the conical singularity,
i.e. the singular set $(\hat r={\cal R},\theta,\phi)$. In the dS geometry this is the
horizon for which the embedding $\Sigma\hookrightarrow{\rm dS}$ has vanishing
extrinsic curvature.

To get the log-divergent term of the entropy we need $b_4$ in backgrounds with
a conical singularity transverse to $\Sigma$. Here we can use the results of
\cite{DeNardo:1996kp} which provides the relevant heat kernel coefficients on cones.
If we expand them around $\alpha=1$,
only the terms proportional to $(\alpha-1)$ will contribute to \eqref{EntropydS}.
What one finds is, in fact, consistent with naive application of the
results of \cite{Fursaev:1995ef}, which computed, to ${\cal O}\big(\alpha-1\big)$,
the integrals of various
curvature invariants over spaces with a regulated conical singularity in the
singular limit (assuming vanishing extrinsic curvature).

We are now ready to list the contributions for the fields of spins $s=0,1/2$ and $1$.
For a real scalar where $\Delta\phi=(-\square+\xi R)\phi$ we find
(with $\tilde b_{4}=180(4\pi)^2 b_4$)
\begin{align}
\int_{{\cal M}_\alpha}\sqrt{g}\,\tilde b_4^{(0,\xi)}&=
\int_{{\cal M}_\alpha}\sqrt{g}\,\left(R^{\mu\nu\rho\sigma}R_{\mu\nu\rho\sigma}
-R^{\mu\nu}R_{\mu\nu}+{5\over2}(1-6\,\xi)^2 R^2
+6(1-5\,\xi)\square\,R\right)\nonumber\\
\noalign{\vskip.2cm}
&\to 8\pi(1-\alpha)\int_\Sigma\left(P^{\mu\rho}P^{\nu\sigma}
R_{\mu\nu\rho\sigma}-{1\over2}P^{\mu\nu}R_{\mu\nu}+{5\over2}(1-6\,\xi)^2 R
\right).
\end{align}
Here ${\cal M}_{\alpha}$ is the manifold with a conical singularity
transverse to $\Sigma$ and
$P^{\mu \nu}$ is the projector into the space orthogonal to the surface $\Sigma$
(in this case the horizon). In the last step we have only kept the term
proportional to $(\alpha-1)$ to which the term $\square\,R$ gives zero
contribution.\footnote{It is worth remarking that
$\int_{{\cal M}_\alpha}\sqrt{g}E_4
\to 32\pi^2(1-\alpha)\chi(\Sigma)$ where $\chi(\Sigma)$ is the Euler number
of $\Sigma$ normalized such that $\chi(S^2)=2$. Furthermore,
$\int_{{\cal M}_\alpha}\sqrt{g}~C^2\to 0$ for $\Sigma\simeq S^2$.}
For dS we have
\begin{align}
R =& \frac{12}{\mathcal{R}^2}\,,\qquad
P^{\mu \nu} R_{\mu \nu}= \frac{6}{\mathcal{R}^2}\,,\qquad
P^{\mu \lambda} P^{\nu \sigma} R_{\mu \nu \lambda \sigma} =\frac{2}{\mathcal{R}^2}
\end{align}
and the integration over the horizon gives $4\pi \mathcal{R}^2$.
This eventually leads to the following expression for the
entropy of a conformally coupled scalar ($\xi=1/6$)
\begin{align}\label{Sscalar}
S_{\rm dS}^{\rm scalar} = \frac{1}{90}  \ln (\delta/\mathcal{R}) \ ,
\end{align}
where we have replaced the momentum cut-off by a distance cut-off $\delta=1/\Lambda$.
This result for the entropy is in agreement with (\ref{logterm0}).

Similarly for a Dirac fermion with $\Delta\psi=(-\square+{1\over 4}R)\psi$
\begin{align}
\int_{{\cal M}_\alpha}\sqrt{g}\,\tilde b_4^{(1/2)}&=
\int_{{\cal M}_\alpha}\sqrt{g}\,\left(-{7\over2}R^{\mu\nu\rho\sigma}R_{\mu\nu\rho\sigma}
-4\,R^{\mu\nu}R_{\mu\nu}+{5\over2}R^2-6\,\square R\right)\nonumber\\
\noalign{\vskip.2cm}
&\to 8\pi(1-\alpha)\int_\Sigma\left(-{7\over 2}P^{\mu\rho}P^{\nu\sigma}R_{\mu\nu\rho\sigma}
-2\,P^{\mu\nu}R_{\mu\nu}+{5\over2}R\right)\\
\noalign{\vskip.2cm}
&~~=11\cdot 32\,\pi^2\,(1-\alpha)\nonumber
\end{align}
leading to
\begin{equation}\label{SDirac}
S_{\rm dS}^{\rm Dirac}={11\over 90}\log(\delta/{\cal R})\,
\end{equation}
also consistent with (\ref{logterm0}).
Finally the contribution of a gauge boson.
Here $\Delta A_\mu=-\square A_\mu+R_{\mu\nu}A^\nu$ (in the gauge $\nabla^\mu A_\mu=0$) and
one finds with the help of \cite{CD}
\begin{align}\label{b4vector}
\int_{{\cal M}_\alpha}\sqrt{g}\,(\tilde b_4^{(1)}&-2\, \tilde b_4^{(0,0)})\\
&=2\int_{{\cal M}_\alpha}\sqrt{g}\,\tilde b_4^{(0,0)}+
\int_{{\cal M}_\alpha}\sqrt{g}\,(-15\,E_4
+30\,R^{\mu\nu}R_{\mu\nu}-15\, R^2
-30\,\square\,R)\nonumber\\
\noalign{\vskip.2cm}
&\to -62\cdot32\,\pi^2\,(1-\alpha)\nonumber
\end{align}
resulting in the following contribution to the entropy:
\begin{align}\label{Svector}
S_{\rm dS}^{\rm gauge\, boson}={62\over90}\log(\delta/{\cal R})\,.
\end{align}
Note that if we had dropped in the second line of \eqref{b4vector}
the contribution proportional to the four-dimensional Euler-density,\footnote{The
remaining terms of the second integral in the second line of \eqref{b4vector}
give zero contribution.} which reduces
to the the Euler number of the singular surface, we would have
obtained the result \eqref{weakentropy}. We will later discuss the relation between this
term and the contact term introduced by Kabat \cite{Kabat:1995eq}.

Combining \eqref{Sscalar},\,\eqref{SDirac} and \eqref{Svector} we find
\begin{equation}
S_{\rm dS}=4\,a\log(\delta/{\cal R})\,
\end{equation}
with $a$ as in \eqref{acoeff}.

\section{Holography and CFT at strong coupling}

At strong coupling, the duality between a CFT at finite temperature on the
$d$-dimensional hyperbolic manifold and a hyperbolic black hole in $d+1$
dimensions in an asymptotically AdS space-time was studied in
\cite{Emparan:1999gf} (see also \cite{Arav:2012ud}) . Holographic duals to
CFTs on the static patch of dS were studied in \cite{Marolf:2010tg}).
The bulk black hole metric is
\begin{align}
ds^2 = -V(r)\, d\tau^2 + \frac{dr^2}{V(r)}+ r^2 dH^2_{d-1} \ ,
\end{align}
where $V(r) = -1 - \frac{M}{r^{d-2}} + \frac{r^2}{\mathcal{R}^2}$. The radius of
the horizon is
\begin{align}
r_h = \frac{2\pi \mathcal{R}^2}{d \beta}
\left(1+ \sqrt{1+\frac{d(d-2) \beta^2}{4\pi^2 \mathcal{R}^2}} \right) \ ,
\end{align}
where $\beta$ is the inverse Hawking temperature. When $\beta = 2\pi {\cal R}$
we have $r_h = \mathcal{R}$, which corresponds to the case where $M=0$.
This is just a patch of pure AdS, analogous to the Rindler patch of Minkowski space.
The horizon entropy is
\begin{align}
S_{\rm BH} = \frac{V_{d-1}}{4\,G_{d+1}} \ ,
\end{align}
where $V_{d-1}$ is the (formally infinite) hyperbolic volume and $G_{d+1}$ Newton's
constant. In the $d=4$ case the universal logarithmic  divergent term reads
\begin{align}
S_{\rm BH} = \frac{\pi\,\mathcal{R}^3}{2\,G_5} \log(\delta/\mathcal{R}) \ .
\end{align}
Comparing with (\ref{logterm0}) we get the identification
\begin{align}
a = \frac{\pi\,\mathcal{R}^3}{8\,G_5} \ .
\end{align}
The five-dimensional Newton constant is \cite{Gubser:1998vd}
\begin{align}
\frac{1}{16\,\pi\, G_5} = \frac{\pi\, N_c^2}{8{\rm Vol}(M_5) \mathcal{R}^3 } \ ,
\end{align}
where we consider ten-dimensional manifolds of the form  $AdS_5 \times M_5$.
For ${\cal N}=4$ SYM theory $M_5 = S^5$ and the volume of the five-sphere
in these conventions, where the Ricci tensor of $M_5$ is
$R_{\mu}{}^{\nu}=4\,\delta_\mu^\nu$, is
$\pi^3$. Plugging this into our formula for $a$, we find
\begin{align}
a_{\rm strong} = \frac{N_c^2}{4} \ .
\end{align}
This is the correct Euler anomaly of ${\cal N}=4$ $SU(N_c)$ SYM at large $N_c$
and it agrees with \eqref{acoeff} for $n_0:n_{1/2}:n_1=6:4:1$ appropriate
for the ${\cal N}=4$ vector multiplet and with $n_1=N_c^2$.
The anomaly coefficient for ${\cal N}=4$ SYM is independent of the gauge coupling constant.
For other conformal field theories, the ratio between their Euler anomaly
coefficients at strong coupling is governed by the ratio of the
Newton's constant in five dimensions, which in
turn is given by the ratio of the volumes of the compact five-manifolds.

Note that for $\mathcal{N}=4$ SYM
on the open Einstein universe one has the same thermal entropy
at the special temperature $(2\pi \mathcal{R})^{-1}$.
In earlier work \cite{Emparan:1999gf} the entropy at
strong coupling was found to be 3/2 times larger than the entropy at weak coupling.
This was puzzling since as was argued in \cite{Emparan:1999gf}, the two
entropies are expected to agree. We now see that the disagreement resulted from the
use of an apparently incorrect entropy for the free spin-1 field.

\section{Stress-energy tensor on conformally flat space-times}

The above analysis implies a modification to the well known formulas
for the vacuum expectation value of the stress-energy tensor
on conformally flat space-times.

In \cite{Candelas:1978gf} it was argued that the thermal stress tensor on the
open Einstein universe is exact to all orders and contains only two possible terms, of orders $\frac{1}{\beta^4}$ and $\frac{1}{\beta^2}$. The $\frac{1}{\beta^4}$ term cannot be modified without affecting the flat space limit (the Stefan-Boltzman law). It turns out the simplest modification that reproduces (\ref{logterm0}) is just a doubling of the spin-1 contribution to the term of order  $\frac{1}{\beta^2}$ in the stress-energy tensor. Thus, instead of (\ref{weakstress}) for open Einstein, we propose
\begin{align}
\langle T_\mu^\nu \rangle_{\rm oE} = \Big[\frac{\pi^2}{90\beta^4}
\left(n_0+\frac{7}{4}n_{1/2}+2\,n_1+\frac{5\beta^2}{8\pi^2\mathcal{R}^2}
(n_{1/2}+16\, n_{1})\right) + C_1\Big]{\rm diag}(-3,1,1,1) \label{oEstress} ,
\end{align}
where we have included a potential spin-1 dependent constant term $C_1$ whose value does not affect the thermal entropy.

We now use the new expression (\ref{oEstress}) together with the relation (\ref{ee}).
for the stress-energy tensor in the other conformally flat space-times.
For the Rindler space-time (\ref{Rindlermetric}) we will have a modified stress-energy tensor
\begin{align}
\langle T_\mu^\nu \rangle_{\rm Rindler} = \frac{{\cal R}^4}{\xi^4} \left( \langle T_\mu^\nu
\rangle_{\rm oE} -  \langle T_\mu^\nu
\rangle_{\rm anom}\right) \ .
\label{Rindlerstress}
\end{align}
The anomaly dependent contribution to the stress tensor $ \langle T_\mu^\nu
\rangle_{\rm anom}$ is (as noted earlier) temperature independent. It is constructed out of curvature tensors and evaluated on the open Einstein universe. In the spin-1 case one finds the following contribution to the energy density
\begin{align}
\epsilon_{\rm anom} = \frac{11}{240 \pi^2 \mathcal{R}^4}.
\end{align}
We demand that the stress-energy tensor for Rindler should vanish at special temperature $B = \frac{2\pi \mathcal{R}}{\beta} =1$, where it is equivalent to the Minkowski vacuum expectation value. This condition fixes the value of $C_1$ above to be $-\frac{1}{24\pi^2 \mathcal{R}^4}$.
The new results for the spin-1 energy densities on Rindler and open Einstein are therefore
\begin{align}
\epsilon^{(s=1)}_{\rm oE} &= \frac{1}{240 \pi^2 \mathcal{R}^4} (B^4+20B^2-10) \\
\epsilon^{(s=1)}_{\rm Rindler} &= \frac{1}{240 \pi^2 \xi^4} (B^4+20B^2-21).
\label{correct}
\end{align}

In the open Einstein universe we now also have a novel zero temperature spin-1 Casimir energy. Note that the energy density at the special value $B=1$ for the open Einstein universe corresponds to the Casimir energy density on the closed Einstein universe, $R \times S^3$. This value, $\frac{11}{240\pi^2 \mathcal{R}^4}$, still agrees with the literature \cite{Candelas:1978gf}. Essentially the doubling of the $B^2$ term is offset by the constant piece. The energy density for a spin-1 field on Rindler space has been previously calculated in several papers
and reads \cite{CandelasDeutsch,Frolov,Dowker,Allen,Moretti:1995fa}
\begin{align}
\epsilon_{(s=1)} = \frac{1}{240\pi^2 \xi^4} (B^4+10B^2-11)  \  . \label{RindlerE}
\end{align}
We will conclude in the next section with a discussion of the potential source of this discrepancy.

\section{Discussion}

At the numerical level, our analysis indicates that a direct calculation of the
entanglement entropy of gauge fields require
some effective doubling of the degrees of freedom localized near the entangling sphere, or equivalently in the thermal picture, the dS horizon.  Understanding the precise meaning of this requires further study (for a recent proposal see \cite{Yamazaki:2013xva}).

Let us discuss now the relation between our modifications to the stress-energy tensor
and the contact term  introduced by Kabat \cite{Kabat:1995eq}.

Kabat's \cite{Kabat:1995eq}  analysis identifies
the heat kernel of a gauge field on Rindler space-time as $d$ (we will use $d=4$) minimally coupled
scalars, minus a contact term
\begin{align}
g_{\mu \nu} K^{\mu \nu}_{\rm vector}(s,x) = 4 K(s,x)_{\rm scalar}
+ \frac{2}{\xi} \partial_{\xi} (s K(s,x)_{\rm scalar}) \ ,
\label{heat}
\end{align}
the second term being the contact contribution (note that one still has
to subtract two ghost scalar contributions from (\ref{heat})). This term was suggested to arise generically due to the difference between the number of
degrees of freedom in the vector zero modes and the ghost scalar modes \cite{Donnelly:2012st}. For example, the difference between the full Maxwell heat kernel and that of two minimally coupled scalars was found to be given by the Euler characteristic, consistent with our results at the end of Section 2.

In our case, the crucial observation is that the energy density of a massless, \textit{minimally coupled} scalar field on Rindler wedge is \cite{Frolov}
\begin{align}
\epsilon_{(s=0)} =  \frac{1}{480\pi^2 \xi^4} (B^2-1)(B^2+11) \ \label{minimale}.
\end{align}
The old result for the spin-1 energy density (\ref{RindlerE}) is twice this value, just as would be expected in $d=4$, but seems to be missing the additional contribution of the contact term.

To evaluate the contact term (\ref{heat}), we used the zeta function regularization. In this approach the free energy density has the form

\begin{align}
F(x) = -\frac{1}{2} \frac{d\zeta(v,x)}{dv}|_{v=0},
\end{align}
where the zeta function is related to the heat kernel by the Mellin transform
\begin{align}
\zeta(v,x) = \frac{1}{\Gamma(v)} \int^{\infty}_{0} s^{v-1} K(s,x) ds.
\end{align}
We worked in the Feynman gauge which was also used by Kabat. We found that the contribution from the Kabat term indeed doubles the order $B^2$ term in the energy density of two scalars as required in order to get from the energy density (\ref{RindlerE}) to the new one (\ref{correct}). That is,
\begin{align}
\epsilon_K = \frac{1}{24\pi^2 \xi^4}(B^2-1) \ .
\end{align}

Unfortunately, our calculation is not conclusive for two reasons. First, in the zeta function regularization it was argued that the Kabat surface term is gauge dependent \cite{Iellici:1996gv}. On the other hand, in the proper time scheme for regulating the heat kernel, one can choose the regulators in such a way that the contribution appears to be gauge independent \cite{Solodukhin:2012jh}. Secondly, the zeta function regularization does not correctly reproduce the energy of the ``bulk" term (\ref{minimale}), although later technology \cite{Moretti:1997qn} seems to fix this problem. More work is needed to settle these issues.

Finally, it would be interesting to explicitly compute the heat kernel for spin-1 fields at finite temperature on the open Einstein universe. This amounts to considering the heat kernel on the space $S^1 \times H^3$, where the circle has periodicity $\beta$. Clearly in this case one would need to carefully investigate global issues, such as the behavior of zero modes.

\section*{Acknowledgements}

This work is supported in part by the ISF center of excellence. We thank
E. Akhmedov, D. Kabat, and Aron Wall for valuable discussions.


\newpage

\begin{thebibliography} {99}

\bibitem{Casini:2011kv}
  H.~Casini, M.~Huerta and R.~C.~Myers,
  JHEP {\bf 1105}, 036 (2011)
  [arXiv:1102.0440 [hep-th]].

\bibitem{Candelas:1978gf}
  P.~Candelas and J.~S.~Dowker,
  Phys.\ Rev.\ D {\bf 19}, 2902 (1979).

\bibitem{CallanWilczek}
  C.~G.~Callan, Jr. and F.~Wilczek,
  Phys.\ Lett.\ B {\bf 333}, 55 (1994)
  [hep-th/9401072].

\bibitem{Ryu:2006ef}
  S.~Ryu and T.~Takayanagi,
  JHEP {\bf 0608}, 045 (2006)
  [hep-th/0605073].

\bibitem{Dowker:2010bu}
  J.~S.~Dowker,
  arXiv:1009.3854 [hep-th].

\bibitem{Kabat:1995eq}
  D.~N.~Kabat,
  Nucl.\ Phys.\ B {\bf 453}, 281 (1995)
  [hep-th/9503016].

\bibitem{Emparan:1999gf}
  R.~Emparan,
  JHEP {\bf 9906}, 036 (1999)
  [hep-th/9906040].


\bibitem{Brown:1986jy}
  M.~R.~Brown, A.~C.~Ottewill and D.~N.~Page,
  Phys.\ Rev.\ D {\bf 33}, 2840 (1986).


\bibitem{BrownCassidy}
  L.~S.~Brown and J.~P.~Cassidy,
  Phys.\ Rev.\ D {\bf 16}, 1712 (1977); T.~S.~Bunch and P.~C.~W.~Davies,
  Proc.\ Roy.\ Soc.\ Lond.\ A {\bf 360}, 117 (1978).

\bibitem{CD}
S.~M.~Christensen and M.~J.~Duff,
Nucl.\ Phys.\ B {\bf 154}, 301 (1979).

\bibitem{DeNardo:1996kp}
  L.~De Nardo, D.~V.~Fursaev and G.~Miele,
  Class.\ Quant.\ Grav.\  {\bf 14}, 1059 (1997)
  [hep-th/9610011].


\bibitem{Fursaev:1995ef}
  D.~V.~Fursaev and S.~N.~Solodukhin,
  Phys.\ Rev.\ D {\bf 52}, 2133 (1995)
  [hep-th/9501127].


\bibitem{Arav:2012ud}
  I.~Arav and Y.~Oz,
  JHEP {\bf 1211}, 014 (2012)
  [arXiv:1206.5936 [hep-th]].

\bibitem{Marolf:2010tg}
  D.~Marolf, M.~Rangamani and M.~Van Raamsdonk,
  Class.\ Quant.\ Grav.\  {\bf 28}, 105015 (2011)
  [arXiv:1007.3996 [hep-th]].

\bibitem{Gubser:1998vd}
  S.~S.~Gubser,
  Phys.\ Rev.\ D {\bf 59}, 025006 (1999)
  [hep-th/9807164].

\bibitem{CandelasDeutsch}
P.~Candelas and D.~Deutsch,
  Proc.\ Roy.\ Soc.\ Lond.\ A {\bf 354}, 79 (1977)

\bibitem{Frolov}
V.~P.~Frolov and E.~M.~Serebryanyi,
Phys.\ Rev.\ D {\bf 35}, 3779 (1987);

  \bibitem{Dowker}
J.~S.~Dowker,
Phys.\ Rev.\ D {\bf 36}, 3742 (1987).

\bibitem{Allen}
B.~Allen, J.~G.~McLaughlin, and A.~C.~Ottewill, Phys.\ Rev.\ D {\bf 45}, 4486 (1992).

\bibitem{Moretti:1995fa}
  V.~Moretti and L.~Vanzo,
  Phys.\ Lett.\ B {\bf 375}, 54 (1996)
  [hep-th/9507139].

\bibitem{Yamazaki:2013xva}
  M.~Yamazaki,
  Europhys.\ Lett.\  {\bf 103}, 21002 (2013)
  [arXiv:1304.0762 [hep-th]].

\bibitem{Donnelly:2012st}
  W.~Donnelly and A.~C.~Wall,
  Phys.\ Rev.\ D {\bf 86}, 064042 (2012)
  [arXiv:1206.5831 [hep-th]].

 \bibitem{Iellici:1996gv}
  D.~Iellici and V.~Moretti,
  Phys.\ Rev.\ D {\bf 54}, 7459 (1996)
  [hep-th/9607015].

\bibitem{Solodukhin:2012jh}
  S.~N.~Solodukhin,
  JHEP {\bf 1212}, 036 (2012)
  [arXiv:1209.2677 [hep-th]].

\bibitem{Moretti:1997qn}
  V.~Moretti,
  Phys.\ Rev.\ D {\bf 56}, 7797 (1997)
  [hep-th/9705060].

\end{thebibliography}
\end{document}